\documentclass[sigconf]{acmart}

\AtBeginDocument{%
  }

\newcommand\blfootnote[1]{%
  \begingroup\renewcommand\thefootnote{}\footnote{#1}\addtocounter{footnote}{-1}\endgroup}

\newcommand{\tslash}{/\allowbreak}

\copyrightyear{2026}
\acmYear{2026}
\setcopyright{cc}
\setcctype{by}
\acmConference[ICCAD '26]{IEEE/ACM International Conference on Computer-Aided Design}{November 08--12, 2026}{San Jose, CA, USA}
\acmBooktitle{IEEE/ACM International Conference on Computer-Aided Design (ICCAD '26), November 08--12, 2026, San Jose, CA, USA}
\acmDOI{10.1145/3831252.3833946}
\acmISBN{979-8-4007-2873-0/2026/11}

\usepackage{wasysym} 
\usepackage{booktabs}
\usepackage{makecell}
\usepackage{multirow}
\usepackage{svg}
\usepackage{placeins}
\renewcommand{\FloatBarrier}{}
\usepackage{listings}
\usepackage{xcolor}

\definecolor{codebg}{HTML}{F5F5F5}
\definecolor{codeframe}{HTML}{CCCCCC}
\definecolor{keyword}{HTML}{0060A8}
\definecolor{string}{HTML}{A31515}
\definecolor{comment}{HTML}{6A737D}

\lstdefinestyle{taidl}{
  language=Python,
  basicstyle=\ttfamily\scriptsize,
  keywordstyle=\color{keyword}\bfseries,
  stringstyle=\color{string},
  commentstyle=\color{comment}\itshape,
  backgroundcolor=\color{codebg},
  frame=single,
  rulecolor=\color{codeframe},
  framesep=3pt,
  xleftmargin=6pt,
  xrightmargin=6pt,
  aboveskip=2pt,
  belowskip=1pt,
  breaklines=true,
  columns=fullflexible,
  keepspaces=true,
  showstringspaces=false,
  tabsize=2,
  morekeywords={add_data_model, add_instruction, add_semantics,
                convert, dot, reshape, transpose},
  literate=
    {<-}{{$\leftarrow$}}2
    {->}{{$\rightarrow$}}2
    {\%A}{{{\color{keyword}\%A}}}2
    {\%B}{{{\color{keyword}\%B}}}2
    {\%C}{{{\color{keyword}\%C}}}2,
  captionpos=b,
}

\begin{document}

\title{TensorLift: Automatic Extraction of Tensor-Level ISA Semantics from Accelerator RTL via MLIR Semantic Lifting}

\author{Ruijie Gao\textsuperscript{1}\authorsep Haoran Jin\textsuperscript{1}\authorsep Jirong Yang\textsuperscript{2*}\lastauthorsep Nathaniel Bleier\textsuperscript{1}}
\affiliation{%
  \institution{\textsuperscript{1}Computer Science and Engineering, University of Michigan, Ann Arbor, MI\\
               \textsuperscript{2}Computer Science, University of Texas at Austin, Austin, TX\\
               \normalfont\{ruijieg, allenjin, nbleier\}@umich.edu \qquad polarisyjr@utexas.edu}
  \country{}
}

\newcommand{\authorsep}{\quad}
\newcommand{\lastauthorsep}{\quad}
\makeatletter
\let\tl@origmkbibcitation\@mkbibcitation
\def\@mkbibcitation{%
  \begingroup
    \def\textsuperscript##1{}%
    \def\authorsep{\unskip{}, \ignorespaces}%
    \def\lastauthorsep{\unskip{}, and \ignorespaces}%
    \tl@origmkbibcitation
  \endgroup}
\makeatother

\renewcommand{\shortauthors}{Gao et al.}

\begin{abstract}

Most proposed tensor accelerators lack well-documented ISAs and compiler backends, and are exercised only through hand-written kernels covering a handful of operators. Recent work (TAIDL, ACT) shows that a tensor-level ISA specification is enough to generate complete software stacks automatically. Writing that specification, however, remains a manual, expert-driven process.

We present TensorLift, the first end-to-end MLIR-based pipeline that lifts RTL-extracted accelerator semantics to TAIDL-like tensor ISA specifications. Building on prior architecture-level model extraction that yields bit-level IR, an 8-pass MLIR pipeline progressively recovers tensor structure---MAC idioms, saturation semantics, multi-dimensional buffer organizations, and layout transformations---and emits specifications the ACT ecosystem consumes directly.

On Gemmini and VTA, TensorLift captures every hardware instruction the two designs decode across 156 MLIR files, reducing the extracted bit-level MLIR by 24.8\% and 41.2\% overall and by up to 92.9\% on a processing element, with controller modules retaining irreducible control logic. It recovers hardware features the hand-written reference omits---multi-bank DMA configuration, pooling, and im2col---and lifts VTA unmodified. Core compute and data-movement semantics are proven equivalent to the RTL-extracted model by Z3 SMT, the remainder validated against golden simulator data. Fed into ACT, the extracted specification yields a compiler backend at parity with hand-written Gemmini kernels ($1.014\times$ geometric mean), giving an automated path from RTL to a working software stack.
\end{abstract}



\maketitle
\blfootnote{\textsuperscript{*}Work done while at the University of Michigan.}

\section{Introduction}
\label{sec:introduction}

\begin{figure}[t]
  \centering
  \includegraphics[width=\linewidth]{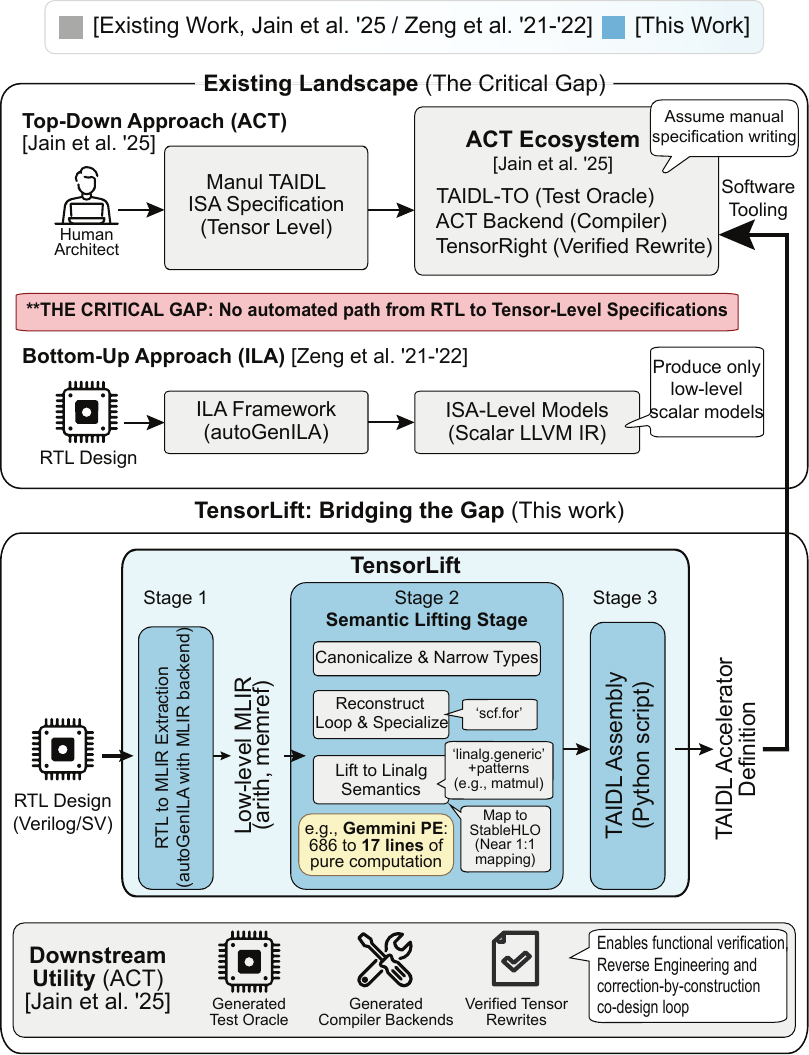}
  \Description{Flow diagram contrasting two existing approaches with this work. A top-down path (the ACT ecosystem) begins with a hand-written, tensor-level TAIDL ISA specification produced by a human architect and yields a test oracle, a compiler backend, and verified rewriting. A bottom-up path (the ILA framework with autoGenILA) begins with an RTL design and yields only scalar-level models in LLVM IR. A red bar between them marks the critical gap: no automated path from RTL to tensor-level specifications. Below, TensorLift spans that gap in three stages, taking Verilog RTL to low-level MLIR, lifting it semantically, and assembling a TAIDL accelerator definition that feeds the same downstream ACT utilities.}
  \caption{Overview of the existing landscape. TensorLift bridges the gap between bottom-up RTL extraction (scalar-level models) and top-down compiler automation (which assumes a hand-written TAIDL specification).}
  \label{fig:overview}
\end{figure}

The rapid growth of deep learning workloads has driven a proliferation of specialized tensor accelerators. Designs such as Gemmini~\cite{gemmini-dac2021}, MAERI~\cite{maeri-asplos2018}, SIGMA~\cite{sigma-hpca2020}, Eyeriss~\cite{eyeriss-jssc2017}, NVDLA~\cite{nvdla}, and FEATHER~\cite{tong2024feather} explore diverse microarchitectural innovations---systolic arrays, reconfigurable interconnects, flexible dataflow engines---for orders-of-magnitude gains in performance and energy efficiency over general-purpose processors. Yet only a handful of vendor-backed platforms---Google's TPU~\cite{tpuv1} with XLA~\cite{xla}, Intel AMX~\cite{amx} with oneDNN~\cite{oneDNN_Contributors_oneAPI_Deep_Neural}, and NVIDIA GPUs with CUDA/cuDNN~\cite{chetlur2014cudnn}---enjoy mature compiler toolchains connecting them to frameworks like PyTorch, JAX~\cite{jax2018github}, and TensorFlow. As Table~\ref{tab:accel-landscape} summarizes, accelerators with end-to-end compiler support number fewer than five, while over a dozen open-source designs lack any usable backend and can be exercised only through hand-written kernel libraries that cover a narrow set of operators and miss cross-layer fusion---leaving most proposed hardware unable to run real ML models end-to-end.

\begin{table}[t]
  \centering
  \scriptsize
  \caption{Software stack readiness of representative open-source tensor/DNN accelerators
  (\CIRCLE{} available, \LEFTcircle{} partial, \Circle{} unavailable). \textbf{TensorLift} targets
  accelerators with open RTL but no ISA specification.}
  \label{tab:accel-landscape}
  \setlength{\tabcolsep}{3pt}
  \begin{tabular*}{\linewidth}{@{\extracolsep{\fill}}lcccc@{}}
    \toprule
    Accelerator
      & \makecell{ISA\\Spec}
      & \makecell{Compiler\\Backend}
      & \makecell{Open\\RTL}
      & \makecell{Test\\Oracle} \\
    \midrule
    MAERI~\cite{maeri-asplos2018}
      & \Circle      & \Circle      & \CIRCLE      & \Circle \\
    SIGMA~\cite{sigma-hpca2020}
      & \Circle      & \Circle      & \CIRCLE      & \Circle \\
    Eyeriss~\cite{eyeriss-jssc2017}
      & \Circle      & \Circle      & \LEFTcircle  & \Circle \\
    MAGNet~\cite{venkatesan2019magnet}
      & \Circle      & \Circle      & \CIRCLE      & \Circle \\
    DNNBuilder~\cite{dnnbuilder-iccad2018}
      & \Circle      & \Circle      & \CIRCLE      & \Circle \\
    FlexASR~\cite{flexasr}
      & \Circle      & \Circle      & \CIRCLE      & \Circle \\
    SPAGHETTI~\cite{spaghetti}
      & \Circle      & \Circle      & \CIRCLE      & \Circle \\
    \midrule
    NVDLA~\cite{nvdla}
      & \LEFTcircle  & \LEFTcircle  & \CIRCLE      & \LEFTcircle \\
    Gemmini~\cite{gemmini-dac2021}
      & \LEFTcircle  & \LEFTcircle  & \CIRCLE      & \LEFTcircle \\
    VTA~\cite{vta}
      & \LEFTcircle  & \LEFTcircle  & \CIRCLE      & \LEFTcircle \\
    DnnWeaver~\cite{dnnweaver-micro2016}
      & \Circle      & \LEFTcircle  & \CIRCLE      & \Circle \\
    \midrule
    Google TPU~\cite{tpuv1}
      & \CIRCLE      & \CIRCLE      & \Circle      & \CIRCLE \\
    Intel AMX~\cite{amx}
      & \CIRCLE      & \CIRCLE      & \Circle      & \CIRCLE \\
    \bottomrule
  \end{tabular*}
\end{table}

The root cause lies upstream of the compiler: the ISA semantics of most accelerators have never been formally defined. Where the CPU ISA is the foundational hardware/software contract, accelerator ISA information is scattered across RTL source, informal documentation, or undocumented designer knowledge~\cite{ila, exo}. Figure~\ref{fig:overview} summarizes the landscape: the top-down ACT ecosystem assumes a hand-written TAIDL specification, while the bottom-up ILA framework produces only scalar-level models, leaving no automated path from RTL to tensor-level semantics.

Recent work closes this gap from the ISA specification downward: TAIDL~\cite{taidl-micro2025} introduced a domain-specific language for defining tensor accelerator ISAs, and ACT~\cite{act-arxiv} showed that compiler backends can be generated automatically from such specifications (\S\ref{subsec:spec-bottleneck}). The prerequisite is critical, though: someone must first write the TAIDL specification by hand, which requires deep understanding of the accelerator's architectural state, instruction encoding, and tensor-level semantics.

This bottleneck is well-documented: writing formal accelerator models has been called ``tedious and error-prone''~\cite{autogenilaiccad} and ``a significant burden for accelerators which may not have the volume or life-span compared to processors''~\cite{autogeniladate}. RTL designs iterate rapidly, while TAIDL and ACT automate everything downstream of the ISA spec but cannot function without one. Prior efforts such as ILA~\cite{ila} and 3LA~\cite{3la} formalized accelerator interfaces but still require hand-authored models, and LLM-based EDA agents~\cite{zhong2024llm4eda, liu2024chipnemo} that depend on machine-readable hardware semantics~\cite{hong2025autocomp, yu2025spec2rtl} cannot operate when such semantics do not exist.

\begin{figure}[t]
  \centering
  \includegraphics[width=\linewidth]{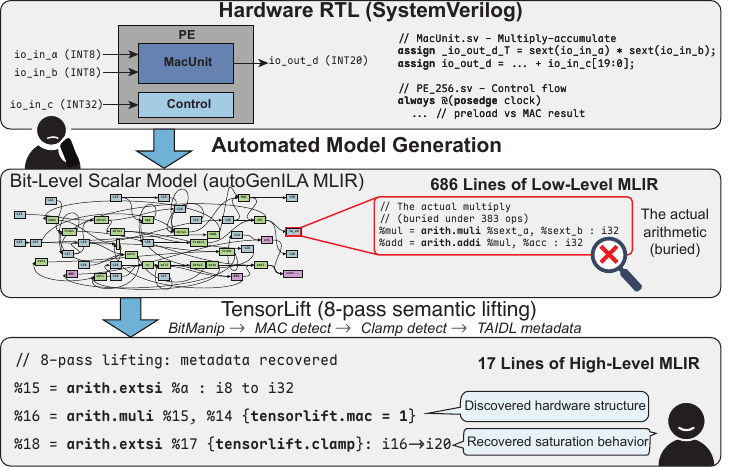}
  \Description{Three stacked code panels for one Gemmini processing element. The top panel shows the SystemVerilog RTL source. The middle panel shows the 686 lines of bit-level MLIR produced by autoGenILA extraction, in which the actual multiply-accumulate arithmetic is buried among sign-extension and bit-packing operations. The bottom panel shows the 17 lines of high-level MLIR that remain after TensorLift lifting, carrying explicit MAC and clamp annotations that name the recovered hardware structure and saturation behavior.}
  \caption{TensorLift's semantic lifting on a Gemmini PE: from RTL source (\textbf{top}) through 686 lines of bit-level MLIR (\textbf{middle}) to 49 lines of lifted MLIR (\textbf{bottom}, 92.9\% reduction).}
  \label{fig:code}
\end{figure}

We present TensorLift, the first end-to-end MLIR-based pipeline that lifts RTL-extracted accelerator semantics to TAIDL-like tensor ISA specifications. Building on prior architecture-level model extraction~\cite{autogeniladate, autogenilaiccad} that produces bit-level MLIR from flattened RTL, TensorLift employs an 8-pass MLIR pipeline that progressively recovers high-level tensor structure---from bit-manipulation canonicalization through idiom detection and loop reconstruction to structured metadata emission---connecting gate-level hardware to the tensor-level semantics tools like TAIDL and ACT require (Figure~\ref{fig:code}).

We evaluate TensorLift on two open-source accelerators: Berkeley's Gemmini (systolic array, 16$\times$16, INT8) and TVM's VTA. Lifting reduces the extracted bit-level MLIR by 24.8\% on Gemmini and 41.2\% on VTA overall, rising to 92.9\% on compute-dominated modules; it discovers hardware features omitted from hand-written references and generalizes across both accelerators without modification. The core compute and data-movement semantics are proven equivalent to the RTL-extracted model via Z3~\cite{z3} SMT, and the remaining lifted operations are validated against golden functional-simulator data. Fed into ACT, the extracted specification yields a compiler backend that matches hand-written Gemmini kernels ($1.014\times$ geometric mean) on workloads including ResNet-50 and MobileNet, with no hand-written specification anywhere in the path.

Filling this first gap in the RTL-to-framework pipeline (Figure~\ref{fig:overview}) removes the step that has required the most manual effort, and completes an automated path from RTL through ACT to an XLA-compatible backend.

In summary, this paper makes the following contributions:

\begin{itemize}
\item We propose the first end-to-end MLIR-based pipeline that lifts RTL-extracted accelerator semantics to TAIDL-like tensor ISA specifications, addressing the manual specification bottleneck that blocks adoption of automated compiler generation tools such as TAIDL and ACT.
\item We design an 8-pass MLIR pipeline that recovers high-level tensor structure from bit-level arithmetic in four phases: canonicalization, idiom detection, loop reconstruction, and metadata emission.
\item We evaluate TensorLift on Gemmini and VTA, reducing extracted bit-level MLIR by 24.8\% and 41.2\% overall and by up to 92.9\% on compute modules, discovering hardware features absent from hand-written references, and formally verifying the core compute and data-movement semantics via Z3 SMT proofs. The same pipeline generalizes across both accelerators without modification.
\item We complete the path from RTL to ML framework, contributing HLO frontend support, convolution pattern matching, and multi-layer memory allocation to ACT, so the extracted specification compiles end-to-end to binaries at parity with hand-written Gemmini kernels ($1.014\times$ geometric mean) on ResNet-50, MobileNet, and others.
\end{itemize}

\section{Background}
\label{sec:background}

\subsection{Hardware-Software Interfaces in Tensor Accelerators}
\label{subsec:hw-sw-interface}

Tensor accelerators occupy a different point in the design space than scalar or vector processors: rather than executing fine-grained instruction streams over flat register files and cache-coherent memory, they expose \emph{coarse-grained} operations over multi-dimensional tensors and replace implicit caches with \emph{explicitly managed} on-chip memories (scratchpads, accumulator banks).
Gemmini~\cite{gemmini-dac2021}, Berkeley's open-source systolic-array generator, is representative: it attaches to a Rocket core~\cite{asanovic2016rocket} via the Rocket Custom Coprocessor (RoCC) interface~\cite{chipyard} with non-standard RISC-V custom instructions; its programmer-visible state is a row-addressed scratchpad and an accumulator; and its ISA exposes only configuration, DMA data-movement, and matrix-multiply execution instructions---every operation simultaneously specifies \emph{what} to compute, \emph{where} operands live in a private memory hierarchy, and \emph{how} data should be staged.

This shifts substantial work onto the compiler.
ML frameworks feed tensor IR (StableHLO, Linalg, TOSA) into backends such as XLA~\cite{xla}, TVM~\cite{tvm}, or IREE~\cite{iree}, which must tile and fuse loop nests to fit scratchpad capacity, schedule explicit DMA transfers, allocate scratchpad regions, and orchestrate double-buffering---all presupposing a precise model of what each instruction does to which piece of tensor state.
Lowering a single XLA \texttt{dot} $C{=}A{\cdot}B$ beyond Gemmini's $16{\times}16$ tile, for instance, requires emitting \texttt{config\_ex}\tslash\texttt{config\_ld}\tslash\texttt{config\_st} to set dataflow and stride parameters, \texttt{mvin}\tslash\texttt{mvin2} to stage tiles of $A$ and $B$, \texttt{preload} followed by \texttt{compute\_preloaded}\allowbreak{} (or \texttt{compute\_accumulated}) to fire the systolic array per output tile, and \texttt{mvout} to drain results---which in turn requires knowing, for every instruction, which tensor tile it touches, which scratchpad region it occupies, and how its output composes with what follows.

Most accelerators fall short here: such a mapping presupposes a \emph{machine-readable} ISA specifying both the binary encoding and the \emph{tensor-level semantics} of each instruction. For Gemmini, semantics live informally in prose documentation, the Chisel RTL, and a hand-written Spike functional model; most open-source accelerators offer only RTL and perhaps a driver, so building a backend today still means \emph{reverse-engineering the RTL by hand}.

\subsection{Automated Compiler Generation and the Specification Bottleneck}
\label{subsec:spec-bottleneck}

The main barrier to end-to-end software support is thus not only backend engineering effort but the lack of a precise, machine-readable Instruction Set Architecture (ISA) specification. TAIDL~\cite{taidl-micro2025} (Tensor Accelerator ISA Definition Language) addresses this with a domain-specific language for describing accelerator data models and instruction semantics at the tensor level, expressing instruction behavior with operators close to the XLA High Level Optimizer (XLA-HLO)~\cite{xla}---\texttt{reshape}, \texttt{transpose}, \texttt{convert}, \texttt{dot}, \texttt{add}. This abstraction is high enough to capture programmer-visible tensor semantics compactly, yet close enough to tensor compiler intermediate representations to support downstream automation, so a representative specification reads less like microarchitectural pseudocode than like a tensor transformation:

\begin{figure}[h]
\begin{lstlisting}[style=taidl, caption={Simplified TAIDL specification for a Gemmini multiply-accumulate instruction (adapted from~\cite{taidl-ae-micro2025}); the full specification additionally includes dataflow routing and state constraint logic.}, label={lst:taidl-example}]
# Data model: scratchpad and accumulator
acc.add_data_model("sp", "1024*DIM", f"{DIM}xs8")
acc.add_data_model("bias", f"{DIM}", f"{DIM}xs32")

# Instruction with register operands
instr = acc.add_instruction("matmul8_compute",
            ["rs1", "rs2"])
instr.add_semantics(f"""
  %A.8:  {DIM}x{DIM}xs8  <- sp[@a.rs1:..., 0:{DIM}];
  %B.8:  {DIM}x{DIM}xs8  <- sp[@a.rs2:..., 0:{DIM}];
  %D.8:  {DIM}x{DIM}xs8  <- sp[@s.rs1:..., 0:{DIM}];
  %A.32: {DIM}x{DIM}xs32 = convert(%A.8);
  %B.32: {DIM}x{DIM}xs32 = convert(%B.8);
  %D.32: {DIM}x{DIM}xs32 = convert(%D.8);
  %dot:  {DIM}x{DIM}xs32 = dot(%A.32, %B.32),
         lhs_contracting_dims={{1}},
         rhs_contracting_dims={{0}};
  %C.32: {DIM}x{DIM}xs32 = add(%dot, %D.32);
  %C.cl: {DIM}x{DIM}xs32 = clamp(-128, %C.32, 127);
  %C.8:  {DIM}x{DIM}xs8  = convert(%C.cl);
  %C.8 -> sp[@s.rs2, 0];
""")
\end{lstlisting}
\Description{A simplified TAIDL specification for a Gemmini multiply-accumulate instruction, written in Python. It declares two data models, a scratchpad and a bias buffer, then adds an instruction taking two register operands whose semantics read three tile operands from the scratchpad, convert them from 8-bit to 32-bit integers, take a dot product with the stated contracting dimensions, add the accumulator tile, clamp the result to the signed 8-bit range, convert back, and write it to the scratchpad.}
\end{figure}

\noindent The author must specify the tensor buffers the accelerator exposes, the instruction operands, and the exact tensor-level effect of the instruction---but not the underlying Register-Transfer Level (RTL) control or datapath implementation. Published examples~\cite{taidl-ae-micro2025} model even Intel Advanced Matrix Extensions (Intel AMX)~\cite{amx} \texttt{tdpbusd} in the same style, composing layout transformations (\texttt{reshape}, \texttt{transpose}), type conversion, tensor contraction, accumulation, and write-back.

Built on such specifications, ACT~\cite{act-arxiv} (Accelerator Compiler Toolkit) automates everything after the ISA description, generating compiler backends by combining equality saturation for instruction selection with constraint-programming-based memory allocation---an ISA-driven flow in which much of the backend logic is synthesized rather than hand-written. This is especially attractive for tensor accelerators, whose parameterized tensor tiles and explicitly managed on-chip memories are difficult to support with conventional backend-generation techniques.

The automation boundary is also the key limitation: TAIDL and ACT automate everything downstream of the ISA specification but not its creation---someone must still write the TAIDL model by hand. This front-end burden is the gap TensorLift addresses.

\subsection{RTL Extraction and MLIR-Based Semantic Lifting}
\label{subsec:rtl-extraction}

The natural baseline for recovering ISA semantics from hardware is architecture-level model extraction from RTL, and the relevant starting point is autoGenILA~\cite{autogenilaiccad, autogeniladate}: it identifies the architectural state variables (ASVs) that persist across instructions, then derives each instruction's state update functions directly from RTL, removing the need to hand-write an Instruction-Level Abstraction (ILA)~\cite{ila} model and yielding an interface consistent with the hardware by construction.
That is its role in TensorLift: the extraction backend that converts accelerator RTL into per-instruction semantic functions.

These extracted models, however, are still too low-level for tensor compiler generation: autoGenILA captures hardware behavior faithfully, but its output is a flattened, bit-level scalar model in which tensor intent is obscured by the structure of the datapath~\cite{autogenilaiccad, autogeniladate}.
As Figure~\ref{fig:code} illustrates, a single Gemmini processing element may initially appear as hundreds of primitive operations---dominated by sign-extension chains, truncations, bitwise composition, and scalar arithmetic---before later passes recover the underlying \texttt{dot\,+\,add\,+\,clamp} semantics.
Such an output suits bit-accurate reasoning but not TAIDL~\cite{taidl-micro2025}, which requires tensor-visible semantics: buffer reads and writes, structured layout transformations, multiply-accumulate patterns, reductions, and saturation behavior.
autoGenILA thus solves the \emph{extraction} problem, but not the \emph{semantic abstraction} problem.

MLIR (Multi-Level Intermediate Representation) is what makes the missing step practical.
MLIR offers no hardware-specific frontend that would help here. What it offers is multi-level, progressive abstraction within one compiler framework~\cite{mlir-cgo2021}, together with the right semantic stepping stones between bit-level arithmetic and tensor-level intent: \texttt{arith} for scalar operations, \texttt{memref} for explicit memory accesses, \texttt{scf} for structured control flow, and \texttt{linalg} for tensor computations~\cite{mlir-pattern-rewriting, mlir-partial-lowering, mlir-linalg-dialect}.
TensorLift materializes the extracted semantics as low-level \texttt{arith}/\texttt{memref} MLIR and raises them through structured, rewrite-driven and partially legalized passes~\cite{mlir-cgo2021, mlir-pattern-rewriting} rather than jumping from RTL-level behavior to tensor semantics in one step (Section~\ref{sec:pipeline}).

\section{TensorLift Pipeline}
\label{sec:pipeline}

\begin{figure}[t]
  \centering
  \includegraphics[width=\linewidth]{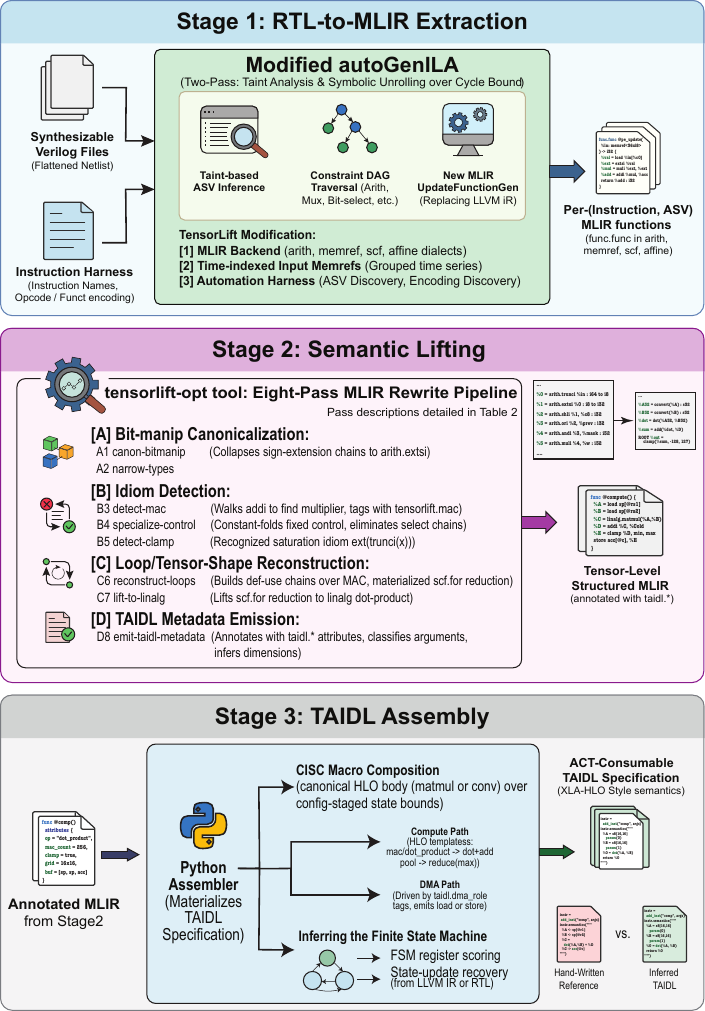}
  \Description{Block diagram of TensorLift as three stages. Stage 1 converts a Verilog or SystemVerilog RTL design into low-level MLIR in the arith and memref dialects. Stage 2, the semantic lifting stage, applies canonicalization and type narrowing, then loop reconstruction and control specialization, then lifting to linalg semantics, and finally emits taidl metadata. Stage 3 assembles that annotated MLIR into a TAIDL accelerator definition through compute and DMA paths together with FSM ordering recovery.}
  \caption{TensorLift's three-stage pipeline: RTL-to-MLIR extraction (Stage~1), eight-pass semantic lifting (Stage~2), and TAIDL assembly with compute/DMA paths and FSM recovery (Stage~3).}
  \label{fig:pipeline}
\end{figure}

TensorLift transforms an unannotated RTL design into a TAIDL specification through three stages (Figures~\ref{fig:overview} and~\ref{fig:pipeline}):
\textbf{Stage~1 (\S\ref{subsec:stage1})} symbolically extracts per-instruction behavior from the RTL into bit-level MLIR;
\textbf{Stage~2 (\S\ref{subsec:stage2})} progressively lifts that bit-level representation to tensor-level semantics via an eight-pass rewrite pipeline;
\textbf{Stage~3 (\S\ref{subsec:stage3})} assembles the lifted MLIR into a TAIDL specification with XLA-HLO-style semantics.
The entire pipeline takes a directory of SystemVerilog files and a top-module name as input, and produces a complete TAIDL specification consumable by the ACT ecosystem.

\subsection{Stage 1: RTL-to-MLIR Extraction}
\label{subsec:stage1}

Stage~1 builds on autoGenILA~\cite{autogenilaiccad,autogeniladate}, which symbolically unrolls a Verilog design for each (instruction, architectural state variable) pair into a function mapping current state and inputs to next state---provably bit-equivalent to the RTL by construction.
Its LLVM IR backend, however, destroys two structural properties that tensor-level reasoning requires: conditional updates are lowered into branches and \texttt{phi} nodes, erasing the if/else structure that encodes hardware mode selection, and RTL signal names---which identify whether an input carries activations, weights, or an accumulator---are discarded.

TensorLift replaces that backend with an MLIR code generator preserving both: conditional updates become \texttt{scf.if} regions, memory accesses remain indexed \texttt{memref} loads, and RTL signal names are attached as structured metadata that downstream passes consult to classify arguments semantically.
A second design choice packs the time series of each input signal across unroll cycles into a single \texttt{memref} argument rather than hundreds of independent scalars---the grouping that later lets loop reconstruction (\S\ref{subsec:stage2}) recover MAC chains as indexable reductions.
An automation layer generates flattening scripts, enumerates architectural state variables, and discovers instruction encodings from the RTL decoder, reducing per-design manual input to two items; the original LLVM IR backend is retained unchanged as a verification oracle (\S\ref{sec:evaluation}).

The output is a corpus of MLIR files, one per (instruction, ASV) pair, provably bit-equivalent to the RTL.
As Figure~\ref{fig:code} illustrates, even a single Gemmini PE produces 686 lines at this stage, its core multiply--accumulate buried under hundreds of sign-extension and bit-packing operations.

\subsection{Stage 2: Semantic Lifting}
\label{subsec:stage2}

Stage~2 is an eight-pass MLIR pipeline that progressively recovers tensor-level semantics from the bit-level Stage~1 output.
The passes fall into four phases (Table~\ref{tab:lifting-passes}).

\textbf{Phase~A (canonicalization)} collapses the bit-by-bit sign-extension chains arising from Verilog \texttt{\$signed} contexts and folds redundant width round-trips---the dominant source of code reduction.
\textbf{Phase~B (idiom detection)} recovers multiply--accumulate by tracing additions back through width casts to a multiplier, dead hardware modes by constant-folding the instruction's fixed control inputs, and fixed-point saturation from the \texttt{ext(trunci(x))} idiom.
\textbf{Phase~C (loop reconstruction)} chains MAC-annotated operations into \texttt{scf.for} reductions and tags those matching the canonical dot-product shape.
\textbf{Phase~D (metadata emission)} classifies each \texttt{memref} argument by access pattern, labels scalar arguments as control attributes, and emits the \texttt{taidl.*} attributes Stage~3 consumes.

\begin{table}[t]
\centering
\caption{TensorLift's eight-pass semantic-lifting pipeline; each pass is named with its \texttt{tensorlift-opt} flag.}
\label{tab:lifting-passes}
\setlength{\tabcolsep}{3pt}
\renewcommand{\arraystretch}{1.0}
\scriptsize
\begin{tabular}{@{}l p{5.9cm}@{}}
\toprule
Phase / Pass & What it does \\
\midrule
A1 \texttt{canon-bitmanip}
& Collapses the bit-by-bit sign-extension chains autoGenILA emits for Verilog \texttt{\$signed} contexts into a single \texttt{arith.extsi}; the dominant source of code reduction on PEs. \\
A2 \texttt{narrow-types}
& Folds redundant \texttt{trunci}/\texttt{ext} round trips left after canonicalization, preserving the \texttt{extsi(trunci(\,))} idioms that Pass~B5 recovers as saturation. \\
\addlinespace[1pt]
B3 \texttt{detect-mac}
& Walks each \texttt{addi} back through width casts to a multiplier and recovers its pre-extension inputs; tags the pair \texttt{tensorlift.mac} when operand widths are hardware-realistic, filtering bit-packing artifacts. \\
B4 \texttt{specialize-control}
& Constant-folds loads of the instruction's fixed control inputs (from the descriptor that drove Stage~1), letting canonicalization eliminate the \texttt{select} chains between hardware modes. \\
B5 \texttt{detect-clamp}
& Recognizes the hardware fixed-point saturation idiom \texttt{ext(trunci(x))} and annotates the surviving extension with the recovered clamp range and signedness. \\
\addlinespace[1pt]
C6 \texttt{reconstruct-loops}
& Builds a use--def chain over MAC-annotated additions whose accumulator input is the previous MAC's output; for chains of length~$\geq\!2$, materializes it as an \texttt{scf.for} reduction with one \texttt{iter\_arg}. \\
C7 \texttt{lift-to-linalg}
& Verifies that a reconstructed \texttt{scf.for} matches the canonical dot-product shape (single \texttt{iter\_arg}, two memref loads at the induction variable, multiply--add--yield) and tags it with \texttt{linalg\_op = "dot\_product"}. \\
\addlinespace[1pt]
D8 \texttt{emit-taidl-metadata}
& Classifies each \texttt{memref} argument by its load/store footprint, labels scalar arguments as control attributes, infers grid dimensions from coordinate suffixes in target ASV names, and emits the closed set of \texttt{taidl.*} attributes Stage~3 consumes. \\
\bottomrule
\end{tabular}
\end{table}

Two properties follow from this design.
First, with the exception of loop reconstruction, passes \emph{annotate} rather than rewrite: they attach attributes to existing operations, and downstream passes key off those attributes.
An instruction that fails to match a template simply receives no annotation, and Stage~3 falls back to opaque semantics rather than producing incorrect TAIDL.
Second, the pipeline is accelerator-agnostic: the structural cues it relies on---sign-extension chains, MAC fan-in, fixed-point clamp idioms, accumulator def--use chains---are properties of common RTL synthesis patterns, not of any particular ISA.
The same binary lifts Gemmini's PE and DMA controllers, and VTA's TensorGemm and TensorAlu without modification.

On the running Gemmini PE example, the eight passes reduce 686 lines of bit-level MLIR to 49, of which only 17 encode the computation core \texttt{clamp(dot(\%A,\%B)+\%C)}.

\paragraph{Scope and limitations.}
These cues are artifacts of how RTL synthesizes arithmetic, so they do not depend on array topology: one binary lifts both a systolic mesh (Gemmini) and an adder-tree GEMM engine (VTA).
Extending that reasoning to classes we do not evaluate: a reconfigurable-interconnect design~\cite{maeri-asplos2018, sigma-hpca2020, tong2024feather} still exposes MAC fan-in and clamp idioms in its multiply and reduction nodes, but its routing fabric is switch and credit logic with no numeric output, so Phase~B finds nothing to annotate and the fabric lifts as control state rather than a tensor operation; a sparse design splits the same way, its value path retaining MAC structure while the coordinate machinery (index merge, run-length decode, descriptor sequencing) stays at the bit level.
In both cases the pipeline emits opaque semantics instead of incorrect TAIDL: passes only annotate, so an unmatched instruction simply receives no annotation.
Synthesis-optimized RTL is the clearest limitation: retiming, register fusion, and technology mapping can redistribute the sign-extension chains and accumulator def--use edges Phases~A and~C depend on, and we do not claim coverage of designs delivered in that form.

\subsection{Stage 3: TAIDL Assembly}
\label{subsec:stage3}

Stage~3 assembles the annotated MLIR into a complete TAIDL specification (Listing~\ref{lst:taidl-example}), merging the per-(instruction, ASV) functions back into per-instruction groups and dispatching on the recognized tensor operation through two paths.
The \emph{compute path} maps each tensor op to an XLA-HLO template---\texttt{dot\_product} to \texttt{convert\,+\,dot\,+\,add} (with optional \texttt{clamp}), \texttt{pool} to \texttt{reduce(max)}, \texttt{im2col\_matmul} to \texttt{reshape\,+\,dot\,+\,add}---inserting the corresponding HLO operation when the RTL encodes an activation function (e.g., \texttt{maximum(result,\,0)} for ReLU).
The \emph{DMA path} classifies memory-port roles (DRAM vs.\ scratchpad address) from the annotated metadata and emits a \texttt{load} or \texttt{store} body accordingly.

For CISC-style loop macros that orchestrate primitive-instruction sequences behind one invocation (e.g., Gemmini's \texttt{loop\_ws}), TensorLift recovers the primitive vocabulary and loop-bound configuration registers from the controller RTL, then composes the primitive tensor operation over those bounds into a whole-program HLO body.
Ordering constraints---that \texttt{compute\_preloaded} may fire only after \texttt{preload}---are recovered by analyzing FSM register updates across instruction groups and folded in as structured metadata.

\section{Evaluation}
\label{sec:evaluation}

We evaluate TensorLift along four axes: how effectively the pipeline compresses bit-level MLIR into tensor-level semantics (\S\ref{subsec:lifting-effectiveness}); whether the lifted semantics are provably equivalent to the RTL (\S\ref{subsec:correctness}); how the extracted specification compares to a hand-written reference (\S\ref{subsec:completeness}); and whether it can drive end-to-end compiler backend generation (\S\ref{subsec:e2e}).

\subsection{Experimental Setup}
\label{subsec:setup}

We target two open-source tensor accelerators.
\textbf{Gemmini}~\cite{gemmini-dac2021} is a systolic-array accelerator (16$\times$16, INT8 MAC) generated via Chipyard~\cite{chipyard} in the \texttt{GemminiRocketConfig} configuration, exposing three hardware controllers (ExecuteController, LoadController, StoreController) with 11 hardware instructions across configuration, DMA data movement, and matrix execution.
\textbf{VTA}~\cite{vta} is TVM's Versatile Tensor Accelerator in the \texttt{DefaultDe10Config} configuration (16-element GEMM engine, ALU, INT8), with four datapath modules (TensorGemm, TensorAlu, Store, GenVMECmd).
All formal verification uses Z3~\cite{z3} bitvector SMT solving, cycle-level validation the Gemmini Spike functional simulator~\cite{gemmini-spike}, and the hand-written reference is the TAIDL MICRO~2025 artifact~\cite{taidl-ae-micro2025}.

\subsection{Semantic Lifting Effectiveness}
\label{subsec:lifting-effectiveness}

\begin{figure}[t]
  \centering
  \includegraphics[width=\linewidth]{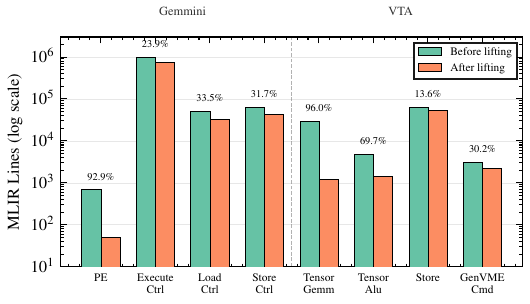}
  \Description{Grouped bar chart with a logarithmic vertical axis of MLIR line counts, comparing before-lifting and after-lifting bars for each module. The four Gemmini modules are PE, ExecuteController, LoadController, and StoreController, reduced by 92.9, 23.9, 33.5, and 31.7 percent. The four VTA modules are TensorGemm, TensorAlu, Store, and GenVMECmd, reduced by 96.0, 69.7, 13.6, and 30.2 percent. The compute-dominated modules, PE and TensorGemm, show by far the largest drops.}
  \caption{MLIR line counts before and after 8-pass lifting, across all Gemmini and VTA modules; percentages indicate reduction.}
  \label{fig:lifting-results}
\end{figure}

\begin{table}[t]
\centering
\caption{Cross-accelerator lifting summary. ``Before'' and ``After'' report total MLIR lines across all per-(instruction, ASV) files in each module.}
\label{tab:lifting-summary}
\setlength{\tabcolsep}{3pt}
\scriptsize
\begin{tabular}{@{}ll r r r r@{}}
\toprule
Accelerator & Module & Files & Before & After & Reduction \\
\midrule
\multirow{4}{*}{Gemmini}
  & PE (TileWithReset) & 1    &     686 &      49 & \textbf{92.9\%} \\
  & ExecuteController  & 80   & 997{,}760 & 759{,}360 & 23.9\% \\
  & LoadController     & 23   &  50{,}004 &  33{,}244 & 33.5\% \\
  & StoreController    & 23   &  63{,}313 &  43{,}269 & 31.7\% \\
\cmidrule{2-6}
  & \textit{Total}     & \textit{127} & \textit{1{,}111{,}763} & \textit{835{,}922} & \textit{24.8\%} \\
\midrule
\multirow{4}{*}{VTA}
  & TensorGemm  &  9 & 29{,}745 &  1{,}183 & \textbf{96.0\%} \\
  & TensorAlu   & 10 &  4{,}832 &  1{,}465 & 69.7\% \\
  & Store       &  5 & 62{,}807 & 54{,}294 & 13.6\% \\
  & GenVMECmd   &  5 &  3{,}160 &  2{,}205 & 30.2\% \\
\cmidrule{2-6}
  & \textit{Total} & \textit{29} & \textit{100{,}544} & \textit{59{,}147} & \textit{41.2\%} \\
\midrule
\textbf{Combined} & & \textbf{156} & \textbf{1{,}212{,}307} & \textbf{895{,}069} & \textbf{26.2\%} \\
\bottomrule
\end{tabular}
\end{table}

Table~\ref{tab:lifting-summary} and Figure~\ref{fig:lifting-results} report the results: TensorLift processes 156 MLIR files comprising 1{,}212{,}307 lines of bit-level MLIR, reducing them to 895{,}069 lines (26.2\% overall).

The reduction varies sharply with module function.
\textbf{Compute-dominated modules} compress most: the Gemmini PE collapses from 686 to 49 lines (92.9\%), its computation core reduced to 17 lines encoding \texttt{clamp(dot(\%A,\%B)+\%C)} once 383 bit-by-bit sign-extension operations from Verilog \texttt{\$signed} contexts fold into single \texttt{arith.extsi} casts under Phase~A; VTA's TensorGemm goes further, its MAC register targets shrinking from 3{,}156 to 8 lines each (99.7\%) and the module from 29{,}745 to 1{,}183 lines (96.0\%).
\textbf{Control-heavy modules} reduce moderately: Gemmini's LoadController (33.5\%) and StoreController (31.7\%) retain address computation, FSM transitions, and instruction muxing---irreducible control logic---while VTA's TensorAlu (69.7\%) falls in between, its opcode muxing across five operations (MIN, MAX, ADD, SHR, SHL) surviving even as the cast chains around each collapse.
\textbf{DMA engines} reduce least: VTA's Store module (13.6\%) and Gemmini's ExecuteController (23.9\%) are dominated by large instruction decoders, multi-level address computation, and deep control logic with little bit-manipulation to canonicalize.
VTA's higher overall reduction (41.2\% vs.\ 24.8\%) reflects which module dominates each total---TensorGemm for VTA, ExecuteController for Gemmini (998K lines across 80 files for 4 instructions $\times$ 20 ASVs).

\paragraph{Pipeline cost.}
Extraction and lifting are a one-time, per-accelerator cost rather than a per-compilation overhead: the extracted MLIR is reused by every subsequent compile.
On Gemmini at DIM${=}16$, end-to-end wall-clock is 1.3\,s for the PE, 6.5\,s for the LoadController, and 7.2\,s for the StoreController; the ExecuteController takes 5{,}250\,s, decomposing into 48\,s flattening, 4{,}540\,s autoGenILA extraction, 619\,s lifting, and 43\,s TAIDL assembly.
Extraction dominates, and its per-PE cost grows super-linearly with array dimension because the symbolic unroll depth scales as $2\cdot\mathrm{DIM}+4$ and the dependency cone deepens with it; both extraction and lifting are embarrassingly parallel across PEs, so wall-clock scales down with available cores.
What this cost replaces is hand-authoring the equivalent TAIDL model, which takes an expert weeks.

\subsection{Correctness Verification}
\label{subsec:correctness}

We verify that TensorLift's lifted semantics are equivalent to the bit-level scalar model they derive from. Since autoGenILA's extraction is provably bit-equivalent to the RTL by construction~\cite{autogenilaiccad, autogeniladate}, equivalence between the lifted MLIR and autoGenILA's output transitively proves \textbf{RTL behavior $\equiv$ TensorLift semantics}.
This machine-checked guarantee covers the core compute and data-movement primitives enumerated in Table~\ref{tab:formal-verification}; the remaining lifted operations---pooling, im2col convolution, and the composed \texttt{mvin}\tslash\texttt{mvout} sequences---are validated differentially against Spike golden data; we do not claim SMT proofs for them.

\begin{table}[t]
\centering
\caption{Formal verification results. Each row is a Z3 bitvector SMT proof establishing equivalence between the lifted MLIR and the bit-level scalar model.}
\label{tab:formal-verification}
\setlength{\tabcolsep}{3pt}
\scriptsize
\begin{tabular}{@{}l l l l@{}}
\toprule
Accelerator & Proof Target & Method & Scope \\
\midrule
\multirow{3}{*}{Gemmini}
  & PE MAC semantics     & Z3 bitvector  & All $2^{60}$ inputs \\
  & WS dataflow mux      & Z3 SMT        & Control specialization \\
  & DMA copy semantics   & Z3 + arrays   & All addresses/values \\
\midrule
VTA & 13 datapath targets & Z3 bitvector  & 4 modules \\
\bottomrule
\end{tabular}
\end{table}

Table~\ref{tab:formal-verification} summarizes the proofs.
For Gemmini we verify the PE's multiply-accumulate core (\texttt{sext(a*b + trunc(c))}) against autoGenILA's LLVM IR output by Z3 bitvector reasoning over all $2^{60}$ input combinations---universal equivalence rather than test vectors---and similarly the weight-stationary dataflow mux (control specialization, Pass~B4) and DMA copy semantics including strided access.
For VTA, 13 targets across all four datapath modules are proven equivalent to the LLVM IR output, covering both compute and control semantics.
Lifting also exposes a structural symmetry: VTA's input and weight index generators produce identical 146-line MLIR after lifting, consistent with the symmetric roles of these buffers in the GEMM engine.
Verification cost is independent of array size: the PE proof discharges in ${\approx}25$\,ms, and by systolic homogeneity that single proof certifies all $\mathrm{DIM}^2$ processing elements, so the cost stays flat as DIM grows.

\paragraph{Guarantees for Stage-3 assembly.}
The proofs above establish Stage-2 lifting; Stage-3 assembly is not itself SMT-proven, and three structural properties bound its risk.
First, HLO templating is a fixed, accelerator-agnostic map from a closed set of Stage-2 annotations to HLO---\texttt{dot\_product} to \texttt{convert\,+\,dot\,+\,add} (the \texttt{add} present only when an accumulator read was recovered), \texttt{mul} to \texttt{multiply} with optional \texttt{clamp}, \texttt{add} to a reduction---with no per-instruction or per-accelerator logic, so dispatch depends only on the recovered annotation.
Second, CISC macro composition (\texttt{loop\_ws}, \texttt{loop\_conv\_ws}) reuses per-primitive MAC semantics already proven equivalent in Stage~2: the macro body is that primitive wrapped in shape plumbing, iterated over a loop nest whose bound \emph{registers} are identified from the controller RTL (the values being runtime configuration fields).
Third, FSM ordering constraints are read off explicit state-comparison logic in the extracted RTL, and every recovered ordering edge corresponds to an instruction that demonstrably writes the guarding control register; the complementary idle-state guard is inferred under a binary-FSM assumption.

As an independent cross-check, all extracted VTA instruction behaviors---GEMM tensor contraction, ALU element-wise operations with five opcodes, Store DMA command generation, and Load burst addressing---align with the operational semantics documented in the VTA specification~\cite{vta} (\S6--9), which the pipeline never reads.

\subsection{Specification Completeness}
\label{subsec:completeness}

TensorLift extracts TAIDL specifications for every hardware instruction Gemmini's three controllers decode, and no instruction on either accelerator falls back to opaque semantics.
Of Gemmini's 11 decoded instructions, seven carry tensor semantics (\texttt{compute\_preloaded},\allowbreak{} \texttt{compute\_accumulated},\allowbreak{} \texttt{preload} operand staging,\allowbreak{} \texttt{mvin}\tslash\texttt{mvin2}\tslash\texttt{mvin3},\allowbreak{} \texttt{mvout}) and four are pure configuration writes producing no numeric output (\texttt{config\_ex}\tslash\texttt{\_ld}\tslash\texttt{\_st}\tslash\texttt{\_norm}); VTA splits 3/1 the same way across its four instructions.
Control instructions have no tensor semantics to recover in the first place, so the opaque fallback never applies to them; their bit-level state updates are preserved verbatim.
VTA's adder-tree GEMM stresses the templates hardest: its reduction does not match the single dense dot-product shape of Pass~C7, so Stage~3 emits it as a composition of primitives with its value semantics intact.

\paragraph{Comparison with hand-written reference.}
Compared against the TAIDL MICRO~2025 artifact~\cite{taidl-ae-micro2025}, a hand-written Gemmini specification, TensorLift captures strictly richer semantics in three areas:

\begin{itemize}
\item \textbf{Multi-bank DMA configuration.}
The hand-written reference models each DMA load variant (\texttt{mvin\_spad}, \texttt{mvin\_acc}, etc.) with a single \texttt{@a.stride} parameter, implicitly assuming all three load engines share one configuration.
Gemmini's LoadController in fact maintains \emph{three independent banks}, each with its own \texttt{stride}, \texttt{scale}, \texttt{shrink}, \texttt{block\_stride}, and \texttt{pixel\_repeat} registers (5 parameters $\times$ 3 banks), the active bank selected by the \texttt{state\_id} field (\texttt{rs1[4:3]}) of the encoding.
The hand-written TAIDL therefore cannot correctly simulate programs using \texttt{mvin} and \texttt{mvin2} with different strides simultaneously---a real case in tiled matrix multiplication, where A- and B-tiles load from DRAM with different access patterns.
TensorLift sees all three banks automatically because autoGenILA traces every register independently and the assembler groups them by indexed base name (\texttt{strides\_0}, \texttt{strides\_1}, \texttt{strides\_2}).

\item \textbf{Pooling engine semantics.}
{\sloppy TensorLift extracts 12 pooling configuration registers from the StoreController (\texttt{pool\_size},\allowbreak{} \texttt{pool\_stride},\allowbreak{} \texttt{pool\_upad},\allowbreak{} \texttt{pool\_lpad},\allowbreak{} \texttt{pool\_orows},\allowbreak{} \texttt{pool\_ocols}, etc.) and generates \texttt{reduce(max) + clamp} semantics; the hand-written reference does not model the pooling engine at all.\par}

\item \textbf{Im2col hardware support.}
TensorLift extracts 9 im2col output ports from the ExecuteController and generates \texttt{reshape + dot} semantics, capturing the hardware's on-the-fly im2col transformation during convolution.
\end{itemize}

The hand-written reference also includes four entries---\texttt{softmax}, \texttt{GELU}, \texttt{layernorm}, and fused add-with-transpose---that are \emph{not} hardware ISA primitives but software-composed sequences using the host CPU's FPU for operations such as \texttt{exp()} and \texttt{divide()}; no Gemmini RTL module implements them.
TensorLift identifies this hardware/software boundary correctly by extracting only instructions with actual RTL decoder entries.

We do not directly compare backends generated from the two specifications because the MICRO~2025 artifact targets a \emph{functional simulator}: it describes how individual instructions move data between scratchpad and accumulator at per-tile (16$\times$16) granularity---what a simulator needs to replay an instruction trace.
A backend instead needs per-\emph{kernel} granularity that maps an entire HLO \texttt{dot} or convolution onto the hardware, and the artifact's per-instruction patterns are too granular for ACT's instruction selection, which expects CISC-level macro-instructions (\texttt{loop\_ws} for a whole matrix multiply, \texttt{loop\_conv\_ws} for a whole convolution).
TensorLift's specification supplies that macro-level abstraction (\S\ref{subsec:stage3}), enabling the backend evaluated next.

\subsection{End-to-End Compiler Backend Generation}
\label{subsec:e2e}

\begin{figure}[t]
  \centering
  \includegraphics[width=\linewidth]{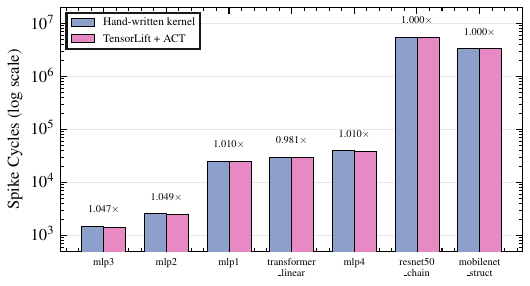}
  \Description{Grouped bar chart with a logarithmic vertical axis of Spike simulator cycle counts, comparing the hand-written kernel against the TensorLift plus ACT backend for seven benchmarks: mlp3, mlp2, mlp1, transformer_linear, mlp4, resnet50_chain, and mobilenet_struct. The paired bars are visually indistinguishable at every benchmark, with per-benchmark speedups of 1.047, 1.049, 1.010, 0.981, 1.010, 1.000, and 1.000 times.}
  \caption{Cycle-level latency of hand-written Gemmini kernels vs.\ the TensorLift+ACT backend on Spike; labels show speedup.}
  \label{fig:latency}
\end{figure}

\FloatBarrier

\begin{table}[t]
\centering
\caption{Latency in Spike simulator cycles; speedup $=$ hand-written\,/\,TensorLift+ACT.}
\label{tab:latency}
\setlength{\tabcolsep}{3pt}
\scriptsize
\begin{tabular}{@{}l r r r@{}}
\toprule
Benchmark & Hand-written & TensorLift+ACT & Speedup \\
\midrule
mlp3               &       1{,}501 &       1{,}433 & $1.047\times$ \\
mlp2               &       2{,}571 &       2{,}451 & $1.049\times$ \\
mlp1               &      25{,}364 &      25{,}123 & $1.010\times$ \\
transformer\_linear &      29{,}732 &      30{,}302 & $0.981\times$ \\
mlp4               &      39{,}742 &      39{,}362 & $1.010\times$ \\
resnet50\_chain     &   5{,}544{,}077 &   5{,}544{,}303 & $1.000\times$ \\
mobilenet\_struct   &   3{,}408{,}414 &   3{,}408{,}581 & $1.000\times$ \\
\midrule
\multicolumn{3}{@{}l}{\textit{Geometric mean}} & $1.014\times$ \\
\bottomrule
\end{tabular}
\end{table}

To validate that the extracted specification suffices for practical use, we feed it into ACT~\cite{act-arxiv} to generate a Gemmini compiler backend and compare the generated code against hand-written kernels on realistic workloads.

\paragraph{Methodology.}
We pass TensorLift's extracted TAIDL specification to ACT, which generates an XLA-compatible backend using equality saturation for instruction selection and constraint programming for scratchpad allocation.
Completing the path to an executable binary required engineering contributions to ACT: HLO frontend support for JAX-produced operations (e.g., \texttt{convolution}, \texttt{reduce\_max}), enode-level preconditions for correct convolution dispatch, and memory allocator support for multi-layer chains.
We reimplement the benchmark suite from gemmini-rocc-tests~\cite{gemmini-rocc-tests}---the Gemmini team's official bare-metal test suite---in JAX~\cite{jax2018github} and compile it through the generated backend, comparing each benchmark's Spike cycle count against the corresponding hand-written C kernel under identical workload parameters (layer shapes, weight-stationary dataflow, activation functions, scaling, batch sizes).
The suite spans MLP, convolutional, and transformer workloads from 2-layer networks (1{,}501 cycles) to full ResNet-50 (5{,}544{,}077 cycles) and MobileNet (3{,}408{,}414 cycles).

\paragraph{Results.}
Table~\ref{tab:latency} and Figure~\ref{fig:latency} report the results.
Across all seven benchmarks, the automatically generated backend lands within $\pm2\%$ of the hand-written kernels ($1.014\times$ geometric mean, per-benchmark range $0.981\times$--$1.049\times$).
The small MLP benchmarks (mlp2, mlp3) run slightly faster ($1.047$--$1.049\times$) from minor tiling differences; the transformer benchmark is the only slowdown ($0.981\times$), from per-tile configuration overhead in the generated sequence; and on the large workloads that dominate real deployment---ResNet-50 (49 conv layers) and MobileNet (52 layers)---the ratio is $1.000\times$.

TensorLift combined with ACT therefore produces a functioning Gemmini backend, at parity with hand-written kernels across the suite, without any hand-written specification.

\section{Related Work}
\label{sec:related}

\paragraph{Formal accelerator interfaces and model extraction.}
Instruction-Level Abstraction (ILA)~\cite{ila} established a software-visible specification framework for accelerators and processors; autoGenILA~\cite{autogenilaiccad, autogeniladate} showed how architecture-level models can be derived automatically from RTL, and 3LA~\cite{3la} how such interfaces support application-level validation.
TensorLift builds on this line but differs in target abstraction: rather than stopping at scalar architectural state updates or validation-oriented interfaces, it lifts extracted semantics to tensor-visible instruction behavior suitable for software-stack generation.

\paragraph{Specification and backend automation.}
TAIDL~\cite{taidl-micro2025} introduced a domain-specific language for expressing tensor-accelerator ISA semantics, and ACT~\cite{act-arxiv} showed that compiler backends can be generated automatically from such specifications.
These efforts automate the downstream path from ISA description to tooling but assume that a tensor-level specification already exists.
TensorLift addresses this missing front-end step by recovering TAIDL-like semantics from RTL-derived low-level models.

\paragraph{MLIR and hardware IR semantics.}
Our work is enabled by MLIR~\cite{mlir-cgo2021} as a progressive abstraction framework and is adjacent to hardware-oriented MLIR efforts such as CIRCT~\cite{circt}; K-CIRCT~\cite{kcirct} and first-class verification dialects for MLIR~\cite{fehr2025verification} further highlight the value of giving hardware and compiler IR dialects explicit semantics.
TensorLift differs in emphasis: rather than defining semantics for an existing IR or hardware dialect, it uses MLIR as the staging ground for lifting extracted hardware behavior to tensor-level ISA descriptions.

\paragraph{Top-down accelerator programming and generation.}
TensorLift should be distinguished from top-down frameworks such as HeteroCL~\cite{heterocl}, Calyx~\cite{calyx}, and Exo~\cite{exo}, which help users design, program, or compile for accelerators from higher-level descriptions; TensorLift instead starts from an existing RTL implementation and reconstructs the software-visible tensor interface.
It has some affinity with semantic lifting and raising in binary analysis~\cite{revng}, but its goal is not generic decompilation to compiler IR---it is recovery of structured tensor semantics consumable by accelerator-specific tooling.

\section{Conclusion}
\label{sec:conclusion}

We presented TensorLift, the first end-to-end pipeline that automatically lifts RTL-extracted accelerator semantics to tensor-level ISA specifications, addressing the manual specification bottleneck that has kept automated compiler generation from reaching most tensor accelerator designs.
Its 8-pass MLIR pipeline---bit-manipulation canonicalization, MAC pattern detection, control-flow specialization, saturation recovery, loop reconstruction, linalg lifting, and structured metadata emission---connects gate-level hardware behavior to the tensor-level semantics languages like TAIDL require.

Evaluated on Gemmini and VTA, TensorLift processes 156 MLIR files totaling over 1.2~million lines of bit-level IR, reducing them by 24.8\% and 41.2\% respectively---up to 92.9\% on compute-dominated modules---and recovering hardware features absent from hand-written references (multi-bank DMA configuration, pooling engine semantics, im2col support).
Core compute and data-movement semantics are proven equivalent to the RTL-extracted model via Z3 SMT, the remainder validated against golden simulator data, and the pipeline generalizes across both architectures unmodified.
Fed into ACT, the extracted specification yields a backend at parity with hand-written kernels ($1.014\times$ geometric mean) up to ResNet-50 and MobileNet, with no hand-written specification anywhere in the path.
TensorLift thereby closes the first gap in the RTL-to-framework pipeline, the step from hardware design to machine-readable ISA semantics, making the full automated path from RTL through TAIDL and ACT to XLA possible.

\setlength{\bibsep}{0pt plus 0.3ex}
\renewcommand{\bibliofont}{\scriptsize}
{\footnotesize
\bibliographystyle{ACM-Reference-Format}
\bibliography{references}

@misc{act-arxiv,
  title = {ACT: Automatically Generating Compiler Backends from Tensor Accelerator ISA Descriptions},
  author = {Jain, Devansh and Pardeshi, Akash and Frigo, Marco and Patel, Krut and Khulbe, Kaustubh and Arora, Jai and Mendis, Charith},
  year = {2025},
  archiveprefix = {arXiv},
  primaryclass = {cs.PL},
  url = {https://arxiv.org/abs/2510.09932},
  doi = {10.48550/arXiv.2510.09932},
  month = oct
}

@inproceedings{taidl-micro2025,
  author = {Jain, Devansh and Frigo, Marco and Arora, Jai and Pardeshi, Akash and Wang, Zhihao and Patel, Krut and Mendis, Charith},
  title = {TAIDL: Tensor Accelerator ISA Definition Language with Auto-generation of Scalable Test Oracles},
  year = {2025},
  isbn = {9798400715730},
  publisher = {Association for Computing Machinery},
  address = {New York, NY, USA},
  url = {https://doi.org/10.1145/3725843.3756075},
  doi = {10.1145/3725843.3756075},
  booktitle = {Proceedings of the 2025 58th IEEE/ACM International Symposium on Microarchitecture},
  pages = {1316–1333},
  numpages = {18},
  series = {MICRO '25},
  month = oct
}

@INPROCEEDINGS{autogenilaiccad,
  author={Zeng, Yu and Huang, Bo-Yuan and Zhang, Hongce and Gupta, Aarti and Malik, Sharad},
  booktitle={2021 IEEE/ACM International Conference On Computer Aided Design (ICCAD)}, 
  title={Generating Architecture-Level Abstractions from RTL Designs for Processors and Accelerators Part I: Determining Architectural State Variables}, 
  year={2021},
  volume={},
  number={},
  pages={1-9},
  doi={10.1109/ICCAD51958.2021.9643584}
}

@INPROCEEDINGS{autogeniladate,
  author={Zeng, Yu and Gupta, Aarti and Malik, Sharad},
  booktitle={2022 Design, Automation \& Test in Europe Conference \& Exhibition (DATE)}, 
  title={Automatic Generation of Architecture-Level Models from RTL Designs for Processors and Accelerators}, 
  year={2022},
  volume={},
  number={},
  pages={460-465},
  doi={10.23919/DATE54114.2022.9774527}
}

@article{ila,
author = {Huang, Bo-Yuan and Zhang, Hongce and Subramanyan, Pramod and Vizel, Yakir and Gupta, Aarti and Malik, Sharad},
title = {Instruction-Level Abstraction (ILA): A Uniform Specification for System-on-Chip (SoC) Verification},
year = {2018},
issue_date = {January 2019},
publisher = {Association for Computing Machinery},
address = {New York, NY, USA},
volume = {24},
number = {1},
issn = {1084-4309},
url = {https://doi.org/10.1145/3282444},
doi = {10.1145/3282444},
journal = {ACM Trans. Des. Autom. Electron. Syst.},
month = dec,
articleno = {10},
numpages = {24}
}

@article{maeri-asplos2018,
  title={Maeri: Enabling flexible dataflow mapping over dnn accelerators via reconfigurable interconnects},
  author={Kwon, Hyoukjun and Samajdar, Ananda and Krishna, Tushar},
  journal={ACM Sigplan Notices},
  volume={53},
  number={2},
  pages={461--475},
  year={2018},
  publisher={ACM New York, NY, USA}
}

@inproceedings{sigma-hpca2020,
  title={Sigma: A sparse and irregular gemm accelerator with flexible interconnects for dnn training},
  author={Qin, Eric and Samajdar, Ananda and Kwon, Hyoukjun and Nadella, Vineet and Srinivasan, Sudarshan and Das, Dipankar and Kaul, Bharat and Krishna, Tushar},
  booktitle={2020 IEEE International Symposium on High Performance Computer Architecture (HPCA)},
  pages={58--70},
  year={2020},
  organization={IEEE}
}

@article{eyeriss-jssc2017,
  title={Eyeriss: An energy-efficient reconfigurable accelerator for deep convolutional neural networks},
  author={Chen, Yu-Hsin and Krishna, Tushar and Emer, Joel S and Sze, Vivienne},
  journal={IEEE journal of solid-state circuits},
  volume={52},
  number={1},
  pages={127--138},
  year={2016},
  publisher={IEEE}
}

@inproceedings{venkatesan2019magnet,
  title={Magnet: A modular accelerator generator for neural networks},
  author={Venkatesan, Rangharajan and Shao, Yakun Sophia and Wang, Miaorong and Clemons, Jason and Dai, Steve and Fojtik, Matthew and Keller, Ben and Klinefelter, Alicia and Pinckney, Nathaniel and Raina, Priyanka and others},
  booktitle={2019 IEEE/ACM International Conference on Computer-Aided Design (ICCAD)},
  pages={1--8},
  year={2019},
  organization={IEEE}
}

@inproceedings{dnnbuilder-iccad2018,
author = {Zhang, Xiaofan and Wang, Junsong and Zhu, Chao and Lin, Yonghua and Xiong, Jinjun and Hwu, Wen-mei and Chen, Deming},
title = {DNNBuilder: an automated tool for building high-performance DNN hardware accelerators for FPGAs},
year = {2018},
isbn = {9781450359504},
publisher = {Association for Computing Machinery},
address = {New York, NY, USA},
url = {https://doi.org/10.1145/3240765.3240801},
doi = {10.1145/3240765.3240801},
booktitle = {Proceedings of the International Conference on Computer-Aided Design},
articleno = {56},
numpages = {8},
location = {San Diego, California},
series = {ICCAD '18}
}

@inproceedings{flexasr,
  title = {A 25mm2 SoC for IoT Devices with 18ms Noise Robust Speech-to-Text Latency
           via Bayesian Speech Denoising and Attention-Based Sequence-to-Sequence
           DNN Speech Recognition in 16nm FinFET},
  author = {Thierry Tambe and En-Yu Yang and Glenn G. Ko and Yuji Chai
            and Coleman Hooper and Marco Donato and Paul N. Whatmough 
            and Alexander M. Rush and David Brooks and Gu-Yeon Wei},
  booktitle = {International Solid-State Circuits Conference (ISSCC)},
  year = {2021}
}

@INPROCEEDINGS{spaghetti,
  author={Hojabr, Reza and Sedaghati, Ali and Sharifian, Amirali and Khonsari, Ahmad and Shriraman, Arrvindh},
  booktitle={2021 IEEE International Symposium on High-Performance Computer Architecture (HPCA)}, 
  title={SPAGHETTI: Streaming Accelerators for Highly Sparse GEMM on FPGAs}, 
  year={2021},
  volume={},
  number={},
  pages={84-96},
  doi={10.1109/HPCA51647.2021.00017}
}

@misc{nvdla,
  author = {NVIDIA Corporation},
  title = {{NVDLA}: Open Source Deep Learning Accelerator},
  year = {2017},
  howpublished = {\url{http://nvdla.org}},
  note = {Accessed: 2026-04-05}
}

@inproceedings{gemmini-dac2021,
  title={Gemmini: Enabling systematic deep-learning architecture evaluation via full-stack integration},
  author={Genc, Hasan and Kim, Seah and Amid, Alon and Haj-Ali, Ameer and Iyer, Vighnesh and Prakash, Pranav and Zhao, Jerry and Grubb, Daniel and Liew, Harrison and Mao, Howard and others},
  booktitle={2021 58th ACM/IEEE Design Automation Conference (DAC)},
  pages={769--774},
  year={2021},
  organization={IEEE}
}

@article{vta,
  title={A hardware--software blueprint for flexible deep learning specialization},
  author={Moreau, Thierry and Chen, Tianqi and Vega, Luis and Roesch, Jared and Yan, Eddie and Zheng, Lianmin and Fromm, Josh and Jiang, Ziheng and Ceze, Luis and Guestrin, Carlos and others},
  journal={IEEE Micro},
  volume={39},
  number={5},
  pages={8--16},
  year={2019},
  publisher={IEEE}
}

@inproceedings{dnnweaver-micro2016,
  title={From high-level deep neural models to FPGAs},
  author={Sharma, Hardik and Park, Jongse and Mahajan, Divya and Amaro, Emmanuel and Kim, Joon Kyung and Shao, Chenkai and Mishra, Asit and Esmaeilzadeh, Hadi},
  booktitle={2016 49th Annual IEEE/ACM international symposium on microarchitecture (MICRO)},
  pages={1--12},
  year={2016},
  organization={IEEE}
}

@inproceedings{tpuv1,
  title={In-datacenter performance analysis of a tensor processing unit},
  author={Jouppi, Norman P and Young, Cliff and Patil, Nishant and Patterson, David and Agrawal, Gaurav and Bajwa, Raminder and Bates, Sarah and Bhatia, Suresh and Boden, Nan and Borchers, Al and others},
  booktitle={Proceedings of the 44th annual international symposium on computer architecture},
  pages={1--12},
  year={2017}
}

@techreport{amx,
  author      = {{Intel Corporation}},
  title       = {Intel Advanced Matrix Extensions (Intel AMX) Technology Brief},
  institution = {Intel Corporation},
  year        = {2023},
  type        = {Technology Brief},
  url         = {https://intel.com}
}

@inproceedings{tong2024feather,
  title={Feather: A reconfigurable accelerator with data reordering support for low-cost on-chip dataflow switching},
  author={Tong, Jianming and Itagi, Anirudh and Chatarasi, Prasanth and Krishna, Tushar},
  booktitle={2024 ACM/IEEE 51st Annual International Symposium on Computer Architecture (ISCA)},
  pages={198--214},
  year={2024},
  organization={IEEE}
}

@misc{xla,
  title	= {XLA : Compiling Machine Learning for Peak Performance},
  author	= {Amit Sabne},
  year	= {2020}
}

@software{oneDNN_Contributors_oneAPI_Deep_Neural,
  author = {{oneDNN Contributors}},
  license = {Apache-2.0},
  title = {{oneAPI Deep Neural Network Library (oneDNN)}},
  url = {https://github.com/uxlfoundation/oneDNN},
  version = {v3.13}
}

@software{jax2018github,
  author = {James Bradbury and Roy Frostig and Peter Hawkins and Matthew James Johnson and Yash Katariya and Chris Leary and Dougal Maclaurin and George Necula and Adam Paszke and Jake Vander{P}las and Skye Wanderman-{M}ilne and Qiao Zhang},
  title = {{JAX}: composable transformations of {P}ython+{N}um{P}y programs},
  url = {http://github.com/jax-ml/jax},
  version = {0.3.13},
  year = {2018},
}

@inproceedings{exo,
  author = {Ikarashi, Yuka and Bernstein, Gilbert Louis and Reinking, Alex and Genc, Hasan and Ragan-Kelley, Jonathan},
  title = {Exocompilation for productive programming of hardware accelerators},
  year = {2022},
  isbn = {9781450392655},
  publisher = {Association for Computing Machinery},
  address = {New York, NY, USA},
  url = {https://doi.org/10.1145/3519939.3523446},
  doi = {10.1145/3519939.3523446},
  booktitle = {Proceedings of the 43rd ACM SIGPLAN International Conference on Programming Language Design and Implementation},
  pages = {703–718},
  numpages = {16},
  location = {San Diego, CA, USA},
  series = {PLDI 2022}
}

@article{3la,
author = {Huang, Bo-Yuan and Lyubomirsky, Steven and Li, Yi and He, Mike and Smith, Gus Henry and Tambe, Thierry and Gaonkar, Akash and Canumalla, Vishal and Cheung, Andrew and Wei, Gu-Yeon and Gupta, Aarti and Tatlock, Zachary and Malik, Sharad},
title = {Application-level Validation of Accelerator Designs Using a Formal Software/Hardware Interface},
year = {2024},
issue_date = {March 2024},
publisher = {Association for Computing Machinery},
address = {New York, NY, USA},
volume = {29},
number = {2},
issn = {1084-4309},
url = {https://doi.org/10.1145/3639051},
doi = {10.1145/3639051},
journal = {ACM Trans. Des. Autom. Electron. Syst.},
month = feb,
articleno = {35},
numpages = {25}
}

@article{chetlur2014cudnn,
  title={cuDNN: Efficient primitives for deep learning},
  author={Chetlur, Sharan and Woolley, Cliff and Vandermersch, Philippe and Cohen, Jonathan and Tran, John and Catanzaro, Bryan and Shelhamer, Evan},
  journal={arXiv preprint arXiv:1410.0759},
  year={2014}
}

@article{liu2024chipnemo,
  title={{ChipNeMo}: Domain-Adapted {LLMs} for Chip Design},
  author={Liu, Mingjie and Ene, Teodor-Dumitru and Kirby, Robert and Cheng, Chris and Pinckney, Nathaniel and Liang, Rongjian and Alben, Jonah and Anand, Himyanshu and Banerjee, Sanmitra and Bayraktaroglu, Ismet and others},
  journal={arXiv preprint arXiv:2311.00176},
  year={2024}
}

@article{zhong2024llm4eda,
  title={{LLM4EDA}: Emerging Progress in Large Language Models for Electronic Design Automation},
  author={Zhong, Ruizhe and Du, Xingbo and Kai, Shixiong and Tang, Zhentao and Xu, Siyuan and Zhen, Hui-Ling and Hao, Jianye and Xu, Qiang and Yuan, Mingxuan and Yan, Junchi},
  journal={arXiv preprint arXiv:2401.12224},
  year={2024}
}

@inproceedings{z3,
  author = {De Moura, Leonardo and Bj\o{}rner, Nikolaj},
  title = {Z3: an efficient SMT solver},
  year = {2008},
  isbn = {3540787992},
  publisher = {Springer-Verlag},
  address = {Berlin, Heidelberg},
  booktitle = {Proceedings of the Theory and Practice of Software, 14th International Conference on Tools and Algorithms for the Construction and Analysis of Systems},
  pages = {337–340},
  numpages = {4},
  location = {Budapest, Hungary},
  series = {TACAS'08/ETAPS'08}
}

@article{hong2025autocomp,
  title={Autocomp: A Powerful and Portable Code Optimizer for Tensor Accelerators},
  author={Hong, Charles and Bhatia, Sahil and Cheung, Alvin and Shao, Yakun Sophia},
  journal={arXiv preprint arXiv:2505.18574},
  year={2025}
}

@inproceedings{yu2025spec2rtl,
  title={Spec2rtl-agent: Automated hardware code generation from complex specifications using llm agent systems},
  author={Yu, Zhongzhi and Liu, Mingjie and Zimmer, Michael and Celine, Yingyan and Liu, Yong and Ren, Haoxing},
  booktitle={2025 IEEE International Conference on LLM-Aided Design (ICLAD)},
  pages={37--43},
  year={2025},
  organization={IEEE}
}

@inproceedings{tvm,
  title={$\{$TVM$\}$: An automated $\{$End-to-End$\}$ optimizing compiler for deep learning},
  author={Chen, Tianqi and Moreau, Thierry and Jiang, Ziheng and Zheng, Lianmin and Yan, Eddie and Shen, Haichen and Cowan, Meghan and Wang, Leyuan and Hu, Yuwei and Ceze, Luis and others},
  booktitle={13th USENIX symposium on operating systems design and implementation (OSDI 18)},
  pages={578--594},
  year={2018}
}

@software{iree,
  author = {{The IREE Authors}},
  license = {Apache-2.0 WITH LLVM-exception},
  month = sep,
  title = {{IREE}},
  url = {https://github.com/iree-org/iree},
  year = {2019}
}

@article{chipyard,
  title={Chipyard: Integrated design, simulation, and implementation framework for custom socs},
  author={Amid, Alon and Biancolin, David and Gonzalez, Abraham and Grubb, Daniel and Karandikar, Sagar and Liew, Harrison and Magyar, Albert and Mao, Howard and Ou, Albert and Pemberton, Nathan and others},
  journal={Ieee Micro},
  volume={40},
  number={4},
  pages={10--21},
  year={2020},
  publisher={IEEE}
}

@software{taidl-ae-micro2025,
  author       = {Devansh Jain and Marco Frigo and Jai Arora and Akash Pardeshi and Zhihao Wang and Krut Patel and Charith Mendis},
  title        = {Artifact of TAIDL: Tensor Accelerator ISA Definition Language with Auto-generation of Scalable Test Oracles},
  year         = {2025},
  month        = aug,
  version      = {1.1.0},
  publisher    = {Zenodo},
  doi          = {10.5281/zenodo.16934755},
  url          = {https://doi.org/10.5281/zenodo.16934755},
  repository   = {https://github.com/act-compiler/taidl-artifact-micro25}
}

@inproceedings{mlir-cgo2021,
  title={MLIR: Scaling compiler infrastructure for domain specific computation},
  author={Lattner, Chris and Amini, Mehdi and Bondhugula, Uday and Cohen, Albert and Davis, Andy and Pienaar, Jacques and Riddle, River and Shpeisman, Tatiana and Vasilache, Nicolas and Zinenko, Oleksandr},
  booktitle={2021 IEEE/ACM International Symposium on Code Generation and Optimization (CGO)},
  pages={2--14},
  year={2021},
  organization={IEEE}
}

@misc{mlir-pattern-rewriting,
  author       = {{LLVM Project}},
  title        = {Pattern Rewriting: Generic {DAG}-to-{DAG} Rewriting},
  howpublished = {\url{https://mlir.llvm.org/docs/PatternRewriter/}},
  year         = {2026},
  note         = {MLIR documentation. Accessed: 2026-04-09}
}

@misc{mlir-partial-lowering,
  author       = {{LLVM Project}},
  title        = {Chapter 5: Partial Lowering to Lower-Level Dialects for Optimization},
  howpublished = {\url{https://mlir.llvm.org/docs/Tutorials/Toy/Ch-5/}},
  year         = {2026},
  note         = {MLIR tutorial documentation. Accessed: 2026-04-09}
}

@misc{mlir-linalg-dialect,
  author       = {{LLVM Project}},
  title        = {Linalg Dialect},
  howpublished = {\url{https://mlir.llvm.org/docs/Dialects/Linalg/}},
  year         = {2026},
  note         = {MLIR documentation. Accessed: 2026-04-09}
}

@misc{gemmini-rocc-tests,
  author       = {{UC Berkeley Architecture Research}},
  title        = {Gemmini RoCC Tests},
  howpublished = {\url{https://github.com/ucb-bar/gemmini-rocc-tests}},
  year         = {2026},
  note         = {Bare-metal test suite for the Gemmini accelerator}
}

@misc{gemmini-spike,
  author       = {{UC Berkeley Architecture Research}},
  title        = {Gemmini Spike Functional Simulator},
  howpublished = {\url{https://github.com/ucb-bar/libgemmini}},
  year         = {2026},
  note         = {Cycle-level functional model for Gemmini}
}

@article{asanovic2016rocket,
  title={The rocket chip generator},
  author={Asanovic, Krste and Avizienis, Rimas and Bachrach, Jonathan and Beamer, Scott and Biancolin, David and Celio, Christopher and Cook, Henry and Dabbelt, Daniel and Hauser, John and Izraelevitz, Adam and others},
  journal={EECS Department, University of California, Berkeley, Tech. Rep. UCB/EECS-2016-17},
  volume={4},
  pages={6--2},
  year={2016}
}

@inproceedings{circt,
  title={MLIR as hardware compiler infrastructure},
  author={Eldridge, Schuyler and Barua, Prithayan and Chapyzhenka, Aliaksei and Izraelevitz, Adam and Koenig, Jack and Lattner, Chris and Lenharth, Andrew and Leontiev, George and Schuiki, Fabian and Sunder, Ram and others},
  booktitle={Workshop on Open-Source EDA Technology (WOSET)},
  volume={3},
  year={2021}
}

@misc{kcirct,
      title={K-CIRCT: A Layered, Composable, and Executable Formal Semantics for CIRCT Hardware IRs}, 
      author={Jianhong Zhao and Jinhui Kang and Yongwang Zhao},
      year={2024},
      eprint={2404.18756},
      archivePrefix={arXiv},
      primaryClass={cs.SE},
      url={https://arxiv.org/abs/2404.18756}, 
}

@article{fehr2025verification,
author = {Fehr, Mathieu and Fan, Yuyou and Pompougnac, Hugo and Regehr, John and Grosser, Tobias},
title = {First-Class Verification Dialects for MLIR},
year = {2025},
issue_date = {June 2025},
publisher = {Association for Computing Machinery},
address = {New York, NY, USA},
volume = {9},
number = {PLDI},
url = {https://doi.org/10.1145/3729309},
doi = {10.1145/3729309},
journal = {Proc. ACM Program. Lang.},
month = jun,
articleno = {206},
numpages = {25}
}

@inproceedings{calyx,
author = {Nigam, Rachit and Thomas, Samuel and Li, Zhijing and Sampson, Adrian},
title = {A compiler infrastructure for accelerator generators},
year = {2021},
isbn = {9781450383172},
publisher = {Association for Computing Machinery},
address = {New York, NY, USA},
url = {https://doi.org/10.1145/3445814.3446712},
doi = {10.1145/3445814.3446712},
booktitle = {Proceedings of the 26th ACM International Conference on Architectural Support for Programming Languages and Operating Systems},
pages = {804–817},
numpages = {14},
location = {Virtual, USA},
series = {ASPLOS '21}
}

@inproceedings{heterocl,
  title={HeteroCL: A multi-paradigm programming infrastructure for software-defined reconfigurable computing},
  author={Lai, Yi-Hsiang and Chi, Yuze and Hu, Yuwei and Wang, Jie and Yu, Cody Hao and Zhou, Yuan and Cong, Jason and Zhang, Zhiru},
  booktitle={Proceedings of the 2019 ACM/SIGDA International Symposium on Field-Programmable Gate Arrays},
  pages={242--251},
  year={2019}
}

@inproceedings{revng,
author = {Di Federico, Alessandro and Payer, Mathias and Agosta, Giovanni},
title = {rev.ng: a unified binary analysis framework to recover CFGs and function boundaries},
year = {2017},
isbn = {9781450352338},
publisher = {Association for Computing Machinery},
address = {New York, NY, USA},
url = {https://doi.org/10.1145/3033019.3033028},
doi = {10.1145/3033019.3033028},
booktitle = {Proceedings of the 26th International Conference on Compiler Construction},
pages = {131–141},
numpages = {11},
location = {Austin, TX, USA},
series = {CC 2017}
}
}

\end{document}